\documentclass[a4]{article}

\font\scap=cmcsc10 \hfuzz=5cm
\font\scap=cmcsc10

\usepackage{graphicx}

\font\tenmsb=msbm10
\font\sevenmsb=msbm7
\font\fivemsb=msbm5
\newfam\msbfam
\textfont\msbfam=\tenmsb
\scriptfont\msbfam=\sevenmsb
\scriptscriptfont\msbfam=\fivemsb
\def\Bbb#1{{\fam\msbfam\relax#1}}

\newcount\eqnumber
\def\neweq{{{(\the\eqnumber)}}\global\advance\eqnumber by 1}
\def\eqdef#1{\eqno\xdef#1{\the\eqnumber}\neweq}
\def\newaeq{{{(\the\eqnumber { a})}}\global\advance\eqnumber by 1}
\def\eqdaf#1{\eqno\xdef#1{\the\eqnumber}\newaeq}
\def\eqdisp#1{\xdef#1{\the\eqnumber}\neweq}
\def\eqdasp#1{\xdef#1{\the\eqnumber}\newaeq}

\newcount\refnumber
\def\newref{{\the\refnumber}\global\advance\refnumber by 1}
\def\refdef#1{{\xdef#1{\the\refnumber}}\newref}

\begin{document}

\centerline{\bf A systematic method for constructing discrete Painlev\'e equations}\centerline{\bf in the degeneration cascade of the E$_8$ group}

\bigskip
\medskip{\scap R. Willox}\quad
{\sl Graduate School of Mathematical Sciences, the University of Tokyo, 3-8-1 Komaba, Meguro-ku, 153-8914 Tokyo, Japan }

\medskip{\scap A. Ramani} and {\scap B. Grammaticos}
\quad{\sl IMNC, Universit\'e Paris VII \& XI, CNRS, UMR 8165, B\^at. 440, 91406 Orsay, France}

\bigskip
{\sl Abstract.} We present a systematic and quite elementary method for constructing discrete Painlev\'e equations in the degeneration cascade for E$_8^{(1)}$. Starting from the invariant for the autonomous limit of the E$_8^{(1)}$ equation one wishes to study, the method relies on choosing simple homographies that will cast this invariant into certain judiciously chosen canonical forms. These new invariants lead to mappings the deautonomisations of which allow us to build up the entire degeneration cascade of the original mapping. We explain the method on three examples, two symmetric mappings and an asymmetric one, and we discuss the link between our results and the known geometric structure of these mappings.
\smallskip



\bigskip
1. {\scap Introduction}
\medskip

Singularities play an essential role in integrability. For continuous systems the study of their singularities gave birth to a most powerful integrability criterion, the famous Painlev\'e property [\refdef\physrep]. The situation for discrete systems is at once similar to that for continuous systems and also more complex. The idea of a singularity in the discrete case is that, starting from generic values, at some point during the iteration of the mapping one suddenly loses a degree of freedom. More specifically, in the case of second-order nonlinear mappings which we shall focus on in this paper, such a situation arises at a value of the variable $x_n$ which is such that its iterate $x_{n+1}$ is independent of $x_{n-1}$. That the  precise iteration step at which this phenomenon occurs actually depends on the initial conditions is what makes the singularity `movable', in analogy to the well-known notion of movable singularities in the continuous case.

One particularly crucial observation led to the formulation of a discrete integrability property which came to be known as {\it singularity confinement} [\refdef\sincon]: by examining a host of discrete systems known to be integrable with spectral methods it was observed that a singularity that appeared at some iteration step, in fact again disappeared after a few more steps. We consider this property to be the discrete analogue of the Painlev\'e property for continuous systems. Though systems that are integrable through spectral methods possess the singularity confinement property, the latter is not a sufficient criterion for integrability. Whereas a general nonintegrable mapping has unconfined singularities (in the sense that a singularity that appears at some iteration never disappears) there exist many examples of systems that have only confined singularities but that are not integrable [\refdef\hiv]. To address this difficulty we introduced in [\refdef\redemp] a method we dubbed {\it full deautonomisation}, which allows to detect the integrable or nonintegrable character of a mapping with confined singularities. 

The absence of unconfined singularities in a given mapping signals the possibility of its regularisation through a succession of blow-ups [\refdef\takenawa,\refdef\dillerfavre]. In [\refdef\redeem] we have used this algebro-geometric method to establish the reliability of our `deautonomisation through singularity confinement' approach.  We remind here that the deautonomisation method consists in extending an autonomous mapping to one where the various, previously constant, coefficients are (appropriately chosen) functions of the independent variable. In a recent paper  [\refdef\cdt], Carstea and co-workers have put this notion of deautonomisation on a rigorous algebro-geometric footing in the case of so-called Quispel-Roberts-Thompson (QRT) [\refdef\qrt] mappings. In particular, they show how to deautonomise a given QRT mapping starting from a particular fibre in the elliptic fibration that is left invariant by the mapping and, as a result, explain how the choice of fibre determines the type of surface the deautonomised mapping will be associated with in  Sakai's classification of discrete Painlev\'e equations [\refdef\sakai].

As a matter of fact, deautonomisation of integrable autonomous mappings, using singularity confinement, is the method through which the vast majority of discrete Painlev\'e equations have been derived. It is, in practice, much faster and easier to apply to a given mapping than a full-blown algebro-geometric analysis. However, both approaches are closely related and often one can be used to shed light on the other. As was shown on a collection of examples in [\refdef\late], not only does the deautonomisation of a mapping through blow-up yield exactly the same conditions on the parameters in the mapping as the singularity confinement criterion  does, these conditions actually turn out to be intimately related to a linear transformation on a particular part of the Picard lattice for the surface obtained from the blow-ups. The part in question is characterised by the fact that its complement is invariant under the automorphism induced on the Picard lattice by the mapping; the automorphism essentially acts as a permutation of the generators for that invariant sublattice.
Quite often, this permutation actually shows up in the singularity analysis of the mapping, as a typical singularity pattern which we call {\it cyclic}. In such a cyclic singularity pattern, a specific pattern of fixed length keeps repeating for all iterations. Note however that such cyclic patterns are not necessarily periodic as some finite values appearing in the repeating pattern may change from one pattern to the next. Strictly speaking, the singularities that are contained in a cyclic pattern are not confined, at least not in the spirit of the original definition as conceived in [\sincon], which demands that once one exits a singularity one should be genuinely home free and not re-enter the same singularity after some further iterations. However, they are most definitely not unconfined singularities either but they should rather be thought of as fixed ones, in the sense that if the initial conditions are not special (i.e. already part of the cycle) then it is impossible to enter such a singularity. Therefore, from an integrability point of view, they are completely innocuous, which is why they are often simply neglected when performing the singularity analysis for a mapping. In [\refdef\halb] Halburd showed how, using the knowledge of confined and cyclic singularities, one can obtain in a simple way the exact degree growth for a given mapping. However, as we showed in [\refdef\expres], when one is interested in a yes or no answer concerning the integrable character of a mapping with both cyclic patterns and confined singularities, the study of the latter does suffice and there is no need to take into account the cyclic singularities. 

Be that as it may, we shall argue here that such cyclic singularity patterns can provide valuable information in the context of the deautonomisation method, as they will often betray the precise identity of the deautonomised mapping in Sakai's classification, even before actually carrying out the deautonomisation. In this paper we are going to study discrete Painlev\'e equations that belong to the degeneration cascade starting from E$_8^{(1)}$ in that classification. Our starting point will be three different equations associated to E$_8^{(1)}$, already obtained in [\refdef\eight]. For each of these equations we start by taking the autonomous limit, thereby obtaining, for each case, a mapping of QRT-type that will play the role of a master mapping. Our guide will be the classification of canonical forms for the QRT mapping which we presented in [\refdef\otonom].
We remind that the QRT invariant, in the symmetric case, has the form
$$K={\alpha_0x_{n-1}^2x_n^2+\beta_0x_{n-1}x_n(x_{n-1}+x_n)+\gamma_0(x_{n-1}^2+x_n^2)+\epsilon_0x_{n-1}x_n+\zeta_0(x_{n-1}+x_n)+\mu_0\over \alpha_1x_{n-1}^2x_n^2+\beta_1x_{n-1}x_n(x_{n-1}+x_n)+\gamma_1(x_{n-1}^2+x_n^2)+\epsilon_1x_{n-1}x_n+\zeta_1(x_{n-1}+x_n)+\mu_1}.\eqdef\qrtinv$$
Eight canonical forms have been obtained in [\otonom]. For completeness we give below the list of the corresponding denominators, $D$, of the QRT invariant (\qrtinv) together with the canonical form of the mapping: 
$$\leqno {\rm(I)} \qquad D=1\qquad\qquad x_{n+1}+x_{n-1}=f(x_n),$$
$$\leqno {\rm(II)} \qquad D=x_{n-1}x_n\qquad\qquad x_{n+1}x_{n-1}=f(x_n),$$
$$\leqno {\rm(III)} \qquad D=x_{n-1}+x_n\qquad\qquad (x_{n+1}+x_{n})(x_{n}+x_{n-1})=f(x_n),$$
$$\leqno {\rm(IV)} \qquad D=x_{n-1}x_n-1\qquad\qquad (x_{n+1}x_{n}-1)(x_{n}x_{n-1}-1)=f(x_n),$$
$$\leqno{\rm(V)}\qquad D=(x_{n-1}+x_n)(x_{n-1}+x_n+2z)\qquad\qquad {(x_{n+1}+x_{n}+2z)(x_{n}+x_{n-1}+2z)\over(x_{n+1}+x_{n})(x_{n}+x_{n-1})}=f(x_n),$$
$$\leqno {\rm(VI)} \qquad D=(x_{n-1}x_n-z^2)(x_{n-1}x_n-1)\qquad\qquad {(x_{n+1}x_{n}-z^2)(x_{n}x_{n-1}-z^2)\over(x_{n+1}x_{n}-1)(x_{n}x_{n-1}-1)}=f(x_n),$$
$$\leqno {\rm(VII)} \qquad D=(x_{n-1}+x_n-z^2)^2-4x_{n-1}x_n\qquad\qquad {(x_{n+1}-x_{n}-z^2)(x_{n-1}-x_{n}-z^2)+x_nz^2\over x_{n+1}-2x_{n}+x_{n-1}-2z^2}=f(x_n),$$
$$\leqno {\rm(VIII)} \qquad D=x_{n-1}^2+x_n^2-(z^2+1/z^2)x_{n-1}x_n+(z^2-1/z^2)^2\quad 
{(x_{n+1}z^2-x_{n})(x_{n-1}z^2-x_{n})-(z^4-1)^2\over(x_{n+1}z^{-2}-x_{n})(x_{n-1}z^{-2}-x_{n})-(z^{-4}-1)^2}=f(x_n).$$
It turns out that in the cases V and VI there exists an alternative way to write the invariant which we have dubbed V$'$ and VI$'$. In this case we have:
$$\leqno{\rm(V')}\qquad D=(x_{n-1}+x_n)^2-z^2\qquad\qquad {(x_{n+1}+x_{n}+z)(x_{n}+x_{n-1}+z)\over(x_{n+1}+x_{n}-z)(x_{n}+x_{n-1}-z)}=f(x_n),$$
$$\leqno {\rm(VI')} \qquad D=x_{n-1}^2+(z+1/z)x_{n-1}x_n+x_n^2\qquad\qquad {(zx_{n+1}+x_{n})(x_{n}+zx_{n-1})\over(x_{n+1}+zx_{n})(zx_{n}+x_{n-1})}=f(x_n).$$
 Implementing homographic transformations of the dependent variable in order to cast the invariant of each master mapping into a judiciously chosen canonical form, we obtain all the possible canonical forms of mappings related to the master mapping. The deautonomisation of these mappings then leads to the discrete Painlev\'e equations in the corresponding degeneration cascade of the original E$_8^{(1)}$ mapping. We study the structure of the singularities for all these mappings and show that apart from the confined singularities which are always there, for certain canonical forms of the invariant and given the right choice of dependent variable, cyclic patterns do also exist. The relation of these patterns to reducible (singular) fibres for the associated invariant and, ultimately, with the method presented in [\cdt] will also be explained.

The equations we shall work with stem from [\eight] where we presented a list of discrete Painlev\'e equations associated to the affine Weyl group E$_8^{(1)}$, written in what we call {\it trihomographic form}. As shown in [\refdef\jmp] the trihomographic form is a very convenient way to represent the various discrete Painlev\'e equations, not limited to those related to E$_8^{(1)}$. The general symmetric form (in the QRT [\qrt] terminology) of the additive trihomographic equation is  
$${x_{n+1}-(z_n+z_{n-1}+k_n)^2\over x_{n+1}-(z_n+z_{n-1}-k_n)^2}{x_{n-1}-(z_n+z_{n+1}+k_n)^2\over x_{n-1}-(z_n+z_{n+1}-k_n)^2}{x_{n}-(2z_n+z_{n-1}+z_{n+1}-k_n)^2\over x_{n}-(2z_n+z_{n-1}+z_{n+1}+k_n)^2}=1,\eqdef\zunu$$
and the asymmetric one 
$${x_{n+1}-(\zeta_n+z_n+k_n)^2\over x_{n+1}-(\zeta_n+z_n-k_n)^2}{x_{n}-(\zeta_n+z_{n+1}+k_n)^2\over x_{n}-(\zeta_n+z_{n+1}-k_n)^2}{y_{n}-(z_n+2\zeta_n+z_{n+1}-k_n)^2\over y_{n}-(z_n+2\zeta_n+z_{n+1}+k_n)^2}=1\eqdaf\zduo$$
$${y_{n}-(\zeta_{n-1}+z_n+\kappa_n)^2\over y_{n}-(\zeta_{n-1}+z_n-\kappa_n)^2}{y_{n-1}-(\zeta_n+z_{n}+\kappa_n)^2\over y_{n-1}-(\zeta_n+z_{n}-\kappa_n)^2}{x_{n}-(\zeta_n+2z_n+\zeta_{n-1}-\kappa_n)^2\over x_{n}-(\zeta_n+2z_n+\zeta_{n-1}+\kappa_n)^2}=1,\eqno(\zduo b)$$
where $z_n, \zeta_n, k_n, \kappa_n$ are functions of the independent variables, the precise form of which is obtained by the application of the singularity confinement criterion. The multiplicative form of the trihomographic equation is obtained from (\zunu) or (\zduo) by replacing the terms $(z_n+z_{n-1}+k_n)^2, $ etc. , by $\sinh^2(z_n+z_{n-1}+k_n), \dots$ [\refdef\multi]. For the elliptic case we have squares of elliptic sines in the left-hand side, i.e. $(z_n+z_{n-1}+k_n)^2, \dots$ should be replaced by ${\rm sn}^2(z_n+z_{n-1}+k_n)$ etc., while in the right-hand side we have a ratio of a product of squares of theta functions,  instead of 1.

Hereafter we shall focus on three particular equations among those obtained in [\eight]. The parameters in these equations are most conveniently expressed using two auxiliary variables. The first one, $t_n$, is related to the independent variable $n$ in a purely secular way by $t_n\equiv\alpha n+\beta$. The second one is a periodic function $\phi_m(n)$ with period $m$, i.e. $\phi_m(n+m)=\phi_m(n)$, given by
 $$ \phi_m(n)=\sum_{l=1}^{m-1} \epsilon_l^{(m)} \exp\left({2i\pi ln\over m}\right).\eqdef\ztre$$
Note that the summation in (\ztre) starts at 1 and not at 0 and thus $\phi_m(n)$ introduces $m-1$ parameters. Two of the equations we are going to study have the symmetric form (\zunu) and parameter dependence
$$z_n=u_{n+1}-u_{n}+u_{n-1},\ k_n=u_{n+2}-u_{n}+u_{n-2} \quad {\rm where}\quad u_n=t_n+\phi_2(n)+\phi_7(n),\eqdef\parsi$$
$$z_n=u_{n},\ k_n=u_{n+1}+u_{n}+u_{n-1} \quad {\rm where}\quad u_n=t_n+\phi_2(n)+\phi_3(n)+\phi_5(n),\eqdef\parsii$$
while the third one has the asymmetric form (\zduo) and parameter dependence:
$$z_n=u_n-u_{n+1}+u_{n-1}-u_{n-2},\ \zeta_n=u_{n+2}-u_n+u_{n-2},\ k_n=u_n,\ \kappa_n=u_{n+2}+u_{n-3}, \quad  u_n=t_n+\phi_8(n).\eqdef\parsiii$$

\bigskip
2. {\scap The symmetric equation with periods 2 \& 7}
\medskip

Before proceeding to the autonomous limit of the additive equation with periods 2 and 7, i.e. (\zunu) with parameter dependence (\parsi), we first study the structure of its singularities already in the nonautonomous form. For this purpose, in fact, it suffices to consider just the secular dependence in $t_n$, neglecting the periodic terms. We have thus the trihomographic expression
$${x_{n+1}-(3t_n-\alpha)^2\over x_{n+1}-(t_n-\alpha)^2}{x_{n-1}-(3t_n+\alpha)^2\over x_{n-1}-(t_n+\alpha)^2}{x_{n}-9t_n^2\over x_{n}-25t_n^2}=1,\eqdef\zcua$$
for which two singularity patterns exist, one of length six $\{x_{n-2}=25t_{n-2}^2, x_{n-1}=(3t_{n-2}-\alpha)^2, x_{n}=(t_{n-2}-4\alpha)^2, x_{n+1}=(t_{n+3}+4\alpha)^2, x_{n+2}=(3t_{n+3}+\alpha)^2, x_{n+3}=25t_{n+3}^2\}$
and one of length four $\{x_{n-1}=9t_{n-1}^2, x_{n}=(t_{n-1}-\alpha)^2, x_{n+1}=(t_{n+2}+\alpha)^2, x_{n+2}=9t_{n+2}^2\}$. 

Next we go to the autonomous limit by taking $\alpha=0$ (and neglecting the periodic terms). Equation (\zcua) now becomes
$${x_{n+1}-9\beta^2\over x_{n+1}-\beta^2}{x_{n-1}-9\beta^2\over x_{n-1}-\beta^2}{x_{n}-9\beta^2\over x_{n}-25\beta^2}=1,\eqdef\zqui$$
and we can put $\beta$ to 1 by a simple rescaling of $x_n$.
Regardless, as shown in [\multi] it suffices to put
$${2\over3}X_n={x_{n}-9\beta^2 \over x_{n}-\beta^2},\eqdef\zsex$$
for (\zcua) to become
$$X_{n+1}X_{n-1}=A{X_n-1\over X_n},\eqdef\zhep$$
for $A=27/4$. (Note that, had we worked with the multiplicative form of the equation, we would have found for $A$ an expression involving $\lambda=e^{\beta}$). The mapping (\zhep) being autonomous and of QRT type it is straightforward to obtain its conserved quantity. We find readily 
$$K={(X_nX_{n-1}+A)(X_n+X_{n-1}-1)\over X_nX_{n-1}},\eqdef\znov$$
where we shall now leave the parameter $A$ essentially free, albeit non-zero.

Next we study the singularity patterns of (\zhep). The singularity of length six for (\zcua) which starts with $x_n=25$ corresponds to a singularity for (\zhep) starting with $X_n=1$. Iterating from this value we find the confined pattern $\{1,0,\infty,\infty,0,1\}$, which fits exactly with the first pattern for (\zcua). Next we turn to the length-four pattern for (\zcua) starting from $x_n=9$, which corresponds to $X_n=0$. Starting with a finite generic value for $X_{n-1}=f$ and from $X_n=0$, we find the following repeating pattern $\{f,0,\infty,\infty,0,f',\infty,f'',0,\infty,\infty,0,\dots\}$ where $f,f',f'',\dots$ are finite values depending on $X_{n-1}$. We remark readily that the pattern $\{f,0,\infty,\infty,0,f',\infty\}$, of length seven, is simply repeated cyclically. Moreover, the confined pattern of length four one would have expected from that of (\zcua), i.e. a pattern $\{0,\infty,\infty,0\}$ in the variable $X_n$, is now embedded in the cyclic one, bracketed by two finite values. 

Deautonomising (\zhep) is straightforward. We assume that $A$ is a function of the independent variable $n$ and require that the confined pattern be still confined in the nonautonomous case. We find the confinement constraint
$$A_{n+2}A_{n-1}=A_{n+1}A_n,\eqdef\zoct$$
which is integrated to $\log A_n=\alpha n+\beta+\gamma(-1)^n$ [\refdef\mulkmt]. This means that the corresponding discrete Painlev\'e equation is associated to the affine Weyl group (A$_1$+A$_1)^{(1)}$. Compared to (\zcua) we lost the period 7 in the coefficients but, on the other hand, it can be checked that the cyclic pattern of length seven persists in the nonautonomous case of (\zhep). Note that in Sakai's classification, a discrete Painlev\'e equation of multiplicative type with (A$_1$+A$_1)^{(1)}$ symmetry is usually associated with a (generalized) Halphen surface of type A$_6^{(1)}$. In the next section we shall explain that this is exactly the intersection type of the reducible singular fibre in the elliptic fibration defined by (\znov) that is responsible for the cyclic singularity pattern of length seven we detected through our singularity analysis.

Next we look for the various ways in which the invariant (\znov) can be made to coincide with one of the 8 canonical forms presented in the introduction. For this we introduce a homographic transformation of the dependent variable (and we add a constant $\kappa$ to the invariant since the latter is at best defined up to an additive constant). We shall not enter into the details of this calculation but give directly the results. Four canonical cases could be identified. For the first two there is no need to add a constant to the invariant.
We find thus that introducing the homography
$$X_n={1\over2}{y_n+z\over y_n-z},\eqdef\zdec$$
(with $z\neq 0$) and giving to $A$ the value $A=-1/4$, we find the invariant
$$K={(y_n^2-z^2)(y_{n-1}^2-z^2)\over (y_n+y_{n-1})(y_n+y_{n-1}-2z)},\eqdef\dunu$$
leading to the mapping
$$\left({y_n+y_{n+1}-2z\over y_n+y_{n+1}}\right)\left({y_n+y_{n-1}-2z\over y_n+y_{n-1}}\right)={y_n-3z\over y_n+z}.\eqdef\dduo$$
On the other hand, the homography
$$X_n={1\over z+1/z}{1-zy_n\over z-y_n},\eqdef\dtre$$
with $A=-1/(z+1/z)^2\neq  -1/4$ leads to the invariant
$$K={(y_n-z)(y_{n-1}-z)(1-zy_n)(1-zy_{n-1})\over (y_ny_{n-1}-z^2)(y_ny_{n-1}-1)},\eqdef\dcua$$
and the mapping
$$\left({y_ny_{n+1}-z^2\over y_ny_{n+1}-1}\right)\left({y_ny_{n-1}-z^2\over y_ny_{n-1}-1}\right)={y_n-z^3\over y_n-1/z}.\eqdef\dqui$$
Deautonomising the mappings (\dduo) and (\dqui) is straightforward and the results have already been presented in [\refdef\maxim]. In short, we find the additive mapping 
$${(y_n+y_{n+1}-z_n-z_{n+1})(y_n+y_{n-1}-z_n-z_{n-1})\over (y_n+y_{n+1})(y_n+y_{n-1})}=
{y_n-z_{n+3}-z_n-z_{n-3} \over y_n-z_{n+3}+z_{n+1}+z_n+z_{n-1}-z_{n-3}},\eqdef\dsex$$
with $z_n=\alpha n+\beta+\phi_7(n)$, from (\dduo), and the multiplicative mapping
$${(y_ny_{n+1}-z_nz_{n+1})(y_ny_{n-1}-z_nz_{n-1})\over (y_ny_{n+1}-1)(y_ny_{n-1}-1)}=
{y_n-z_{n+3}z_nz_{n-3} \over y_n-z_{n+3}z_{n-3}/(z_{n+1}z_nz_{n-1})},\eqdef\dhep$$
with $\log z_n=\alpha n+\beta+\phi_7(n)$, from (\dqui). Both have E$_7^{(1)}$ symmetry and, in particular, it should be noted that the period 2 which was present in the parameters of the original mapping we started from, has been lost. A study of their singularities shows that both equations have two confined patterns of lengths 6 and 4, just like (\zcua), and this even when one neglects the periodic term in $z_n$, keeping only the secular dependence. Thus, in the nonautonomous case the length-7 cyclic pattern has disappeared but a periodic term of period 7 has made its appearance in the coefficients of the equation. 

The interesting thing is the appearance of the length-four singularity. The way singularity confinement was always implemented in the deautonomisation procedure was to look for the confined singularity patterns in the autonomous case and preserve them upon deautonomisation. The existence of cyclic patterns where a shorter pattern--in this case of length four--is bracketed by two finite values shows that there exist situations where one may decide that a pattern is confined where, in reality, it is just a part of a longer, cyclic, one. However it turns out that such a misidentification is innocuous in practice. First, as explained in the introduction, the existence of cyclic singularity patterns is compatible with integrability, and second, as seen here, upon deautonomisation the pattern embedded between two finite values may indeed become confined. 

There are still two other possible homographies we can implement. Obviously, when $A=27/4$, the homography (\zsex) is among the possible ones, leading back to the mapping we started with. Which leaves one final homography which is of the form
$$X_n={z^2+1+1/z^2\over z^2+1/z^2}{y_n+z^3+1/z^3\over y_n+z+1/z}.\eqdef\doct$$
Together with the values $A=(z^2+1+1/z^2)^3/(z^2+1/z^2)^2$ and $\kappa=-(z+1/z)^4/(z^2+1/z^2)$ this yields the invariant
$$K={(y_n+z+1/z)(y_{n-1}+z+1/z)(y_n+z^3+1/z^3)(y_{n-1}+z^3+1/z^3)\over y_n^2+y_{n-1}^2-(z^2+1/z^2)y_ny_{n-1}+(z^2-1/z^2)^2}.\eqdef\dnov$$
The corresponding mapping is
$${(y_n-z^2y_{n+1})(y_n-z^2y_{n-1})-(z^4-1)^2\over (z^2y_n-y_{n+1})(z^2y_n-y_{n-1})-(z^4-1)^2/z^4}={y_n+z^4(z+1/z)\over z^4y_n+z+1/z}.\eqdef\ddec$$
Note that this is the canonical form of the autonomous limit of a multiplicative discrete Painlev\'e equation associated to the affine Weyl group E$_8^{(1)}$. Its deautonomisation has been presented in [\multi].

It is interesting to point out here that elliptic discrete Painlev\'e equations can also be obtained, be it at the price of slightly generalising our starting point to  the autonomous, trihomographic, form 
$${\delta x_{n+1}-\eta\over \mu x_{n+1}-\nu}{\delta x_{n-1}-\eta\over \mu x_{n-1}-\nu}{\delta x_{n}-\eta\over \rho x_{n}-\sigma}=1,\eqdef\wunu$$
where the parameters are expressed in terms of theta functions as follows: $\delta=\theta_0^2(3z)$, $\eta=\theta_1^2(3z)$, $\mu=\theta_0^2(z)$, $\nu=\theta_1^2(z)$, $\rho=\theta_0^2(5z)$, $\sigma=\theta_1^2(5z)$.
(Here we are using the conventions of the monograph of Byrd and Friedman [\refdef\byrd] for the naming of the theta functions). Introducing the homography
$${\delta\sigma-\eta\rho\over\mu\sigma-\nu\rho}X_n={\delta x_n-\eta \over \mu x_n-\nu},\eqdef\wduo$$
we obtain precisely equation (\zhep) with $A=(\mu\sigma-\nu\rho)^3/((\delta\nu-\eta\mu)(\delta\sigma-\eta\rho)^2)$. Thus the analysis presented above also leads to elliptic equations associated to the E$_8^{(1)}$ group. For example, starting from the parameter dependence (\parsi) and introducing
$$\displaylines{a_n=z_n+z_{n-1}+k_n,\ b_n=z_n+z_{n-1}-k_n,\ c_n=z_n+z_{n+1}+k_n,\ d_n=z_n+z_{n+1}-k_n, \hfill\cr\hfill e_n=2z_n+z_{n-1}+z_{n+1}-k_n,\ f_n=2z_n+z_{n-1}+z_{n+1}+k_n,\quad\eqdisp\fduo\cr}$$
we obtain the full, nonautonomous, form of this discrete Painlev\'e equation:
$${\theta_0^2(a_n)x_{n+1}-\theta_1^2(a_n)\over \theta_0^2(b_n)x_{n+1}-\theta_1^2(b_n)}\ {\theta_0^2(c_n)x_{n-1}-\theta_1^2(c_n)\over \theta_0^2(d_n)x_{n-1}-\theta_1^2(d_n)}\ {\theta_0^2(e_n)x_{n}-\theta_1^2(e_n)\over  \theta_0^2(f_n)x_{n}-\theta_1^2(f_n)}= 1.\eqdef\ftre$$ 

The equations presented above are not the only ones we can obtain starting from (\zhep) or, equivalently, (\zqui). Given the form of (\zhep) it is clear that we can also consider an evolution where we skip one out of two indices and thus establish a mapping relating $x_{n+2}, x_n$ and $x_{n-2}$. The simplest way is to solve (\zhep) for $X_n$ and obtain an invariant involving $X_{n+1}$ and $X_{n-1}$. Downshifting the indices we find 
$$K={(X_nX_{n-2}-X_n-A)(X_nX_{n-2}-X_{n-2}-A)\over X_nX_{n-2}-A}.\eqdef\vunu$$
Putting $A=a^2$ and rescaling $X$, so as to absorb one factor of $a$, we obtain from (\vunu) the mapping
$$(X_nX_{n+2}-1)(X_nX_{n-2}-1)={X_n\over a^2X_n-a}.\eqdef\vduo$$
The singularity patterns of (\vduo) can be directly read off from those of (\zhep). We have the confined pattern $\{1/a,\infty,0\}$, its mirror image $\{0,\infty,1/a\}$ and a cyclic one $\{\infty,0,\infty,0,\infty,f,f'\}$,
 of length 7, as expected. In order to deautonomise (\vduo) we write its right-hand side as $X_n/(b_nX_n-a_n)$. Applying the confinement criterion we then find $\log a_n=\alpha n+\beta$ and $b_n=da_n^2$ where $d$ is a free constant [\mulkmt]. Note that in this case the parameters in the mapping only have secular dependence on $n$.
 
It is interesting at this point to introduce the invariant associated to the autonomous limit of (\vduo) for the right-hand side $X_n/(d a^2 X_n-a)$. We find
$$K={a^2dX_n^2X_{n-2}^2-a X_nX_{n-2}(X_n+X_{n-2})+a(X_n+X_{n-2})-a^2d+1\over X_nX_{n-2}-1}\eqdef\aunu$$
In order to obtain double-step equations associated to E$_7^{(1)}$ or E$_8^{(1)}$ we can now either start from (\dduo) and (\dqui), or use the invariant (\aunu). Both approaches are not equivalent however, since working with (\dduo) and (\dqui) is tantamount to setting $d=1$ (and thus to working with equations obtained from (\znov)), which turns out to be a necessary constraint to obtain  E$_7^{(1)}$ related equations, but not for those related to E$_8^{(1)}$.

Working with (\dqui), along the lines set out above, we find the invariant
$$K={(y_{n-2}-z)(y_n-z)\big((z+1/z)(y_{n-2}y_n+1)-2(y_{n-2}+y_n)\big)\over(z^2y_n-y_{n-2})(z^2y_{n-2}-y_n)},\eqdef\vtre$$
and the corresponding mapping
$$\left({z^2y_{n+2}-y_n\over y_{n+2}-z^2y_n}\right)\left({z^2y_{n-2}-y_n\over y_{n-2}-z^2y_n}\right)={(y_n-z)^2(y_n-z^3)\over z(zy_n-1)^3}.\eqdef\vcua$$
We remark that the mapping is given in the form we dubbed (VI$'$) in [\otonom]. Its deautonomisation, albeit in the canonical form (VI), was given in [\refdef\seven], where we have derived E$_7^{(1)}$-associated equations with period-7 periodicity in their parameters. The non-autonomous form of (\vcua) corresponds to equation (20) of [\seven]. An analogous derivation centred around (\dduo) leads to the mapping
$$\left({y_{n+2}-y_n+2z\over y_{n+2}-y_n-2z}\right)\left({y_{n-2}-y_n+2z\over y_{n-2}-y_n-2z}\right)={(y_n-z)^2(y_n-3z)\over (y_n+z)^3}.\eqdef\vqui$$
This is a mapping of canonical form (V$'$). Clearly it can be brought to the canonical (V) form through some elementary transformations: changing the sign of two consecutive $y_n$ out of four we can bring the left hand side to one where the dependent variables have the same sign whereupon a simple translation can absorb the terms proportional to $z$ in the denominator. Again the deautonomisation results can be found in [\seven], equation (11).

Where the explicit presence of the parameter $d$ becomes essential is in the derivation of the double-step equations associated to the E$_8^{(1)}$ group. For this, we start from the invariant (\aunu) and look for the proper homographic transformation bringing the invariant to a canonical E$_8^{(1)}$ form. We focus on the additive case, for which a detailed analysis was presented in [\refdef\ancil]. Without going into the lengthy details we find that the autonomous mapping we obtain is the autonomous limit of the equation 5.2.7 of [\ancil] where the parameter $d$ above and the constant $c$ figuring in 5.2.7 are related through $d=(c-3)(c+1)^3/((c+3)(c-1)^3)$. 
The deautonomisation of this mappings is obviously the one given in [\ancil]. Thus equation (\zcua), which corresponds to case 3.3 in [\ancil], is related, once we consider evolution with double step, to case 5.2.7 of that paper. It goes without saying that similar results could be obtained for multiplicative and elliptic systems.

\bigskip
3. {\scap Invariants, singular fibres and deautonomisations}
\medskip

As is well-known since the seminal works of Veselov [\refdef\veselov] and Tsuda [\refdef\tsuda], a QRT map with invariant (\qrtinv) can be described by a birational automorphism on a rational elliptic surface. This surface, say $X$, is obtained from blowing up ${\Bbb P}^1\times{\Bbb P}^1$ at eight points $(u_i,v_i)$. These are the eight points (taking into account multiplicities) that do not lie on a unique member of the pencil of quadratic curves
$$\displaylines{ (\alpha_0+K\alpha_1)u^2v^2+(\beta_0+K\beta_1)uv(u+v)+(\gamma_0+K\gamma_1)(u^2+v^2)\hfill\cr\hfill+
(\epsilon_0+K\epsilon_1)uv+(\zeta_0+K\zeta_1)(u+v)+(\mu_0+K\mu_1)=0,\quad \eqdisp\antonii\cr}$$
obtained from the invariant (\qrtinv) by taking $x_{n-1}=u$, $x_n=v$. Each member of this family of quadratic curves (\antonii), parametrised by $K\in{\Bbb P}^1$, is invariant under the QRT map at hand. 
What is important here, is that after blow-up, the QRT map is so to speak `lifted' to an automorphism 
on $X$ that preserves the elliptic fibration $\pi: X \to {\Bbb P}^1$ described by (\antonii). For a generic value of $K$, the fibre $\pi^{-1}(K)\in X$ is a genuine elliptic curve (i.e. an irreducible smooth curve of genus one). There exist however a finite number of special values for $K$ for which the corresponding curve is either not smooth simply because it contains a node or a cusp (the so-called multiplicative and additive reductions, respectively) or because it is in fact reducible and therefore  contains singular points.  The fibres $\pi^{-1}(K)$ that correspond to these special values of $K$ are called {\it singular} and are either reducible or irreducible, in the sense explained above. Reducible fibres can be represented graphically by their intersection diagrams, which are actually Dynkin diagrams of ADE-type. For surfaces with a section, which is the case for the QRT-maps, all possible configurations of reducible fibres have been studied in [\refdef\os], resulting in a relatively short list of 74 different surface types, which makes the list rather easy to consult. A full classification, including all possible configurations that involve irreducible singular fibres, and which also distinguishes between multiplicative and additive A$_1^{(1)}$ or A$_2^{(1)}$ fibres (which is not the case in [\os]), can be found in [\refdef\pers] and [\refdef\mir] but the result is a much longer list (almost four times as long!) which is much less accessible than the one in [\os]. In the following we shall mainly concern ourselves with reducible singular fibres for the invariants we wish to treat, as these are easily obtained without resorting to special algorithms (see e.g. Appendix C in [\cdt]) and, whenever necessary, we shall use the lists in [\pers] or [\mir] to supplement our results with information on the irreducible fibres.

Let us consider for example the three point mapping (\zhep) where we take $X_{n-1}$ to be $u_n$ and $X_n$ to be $v_n$ and where $A$ is a non-zero (complex) constant. Note that, rewriting the invariant (\znov) in these new variables, we see readily that $K$ becomes infinite for the four lines $u=0$, $v=0$, $u=\infty$, $v=\infty$ on ${\Bbb P}^1\times{\Bbb P}^1$. Moreover, the mapping (\zhep) has two indefinite points, $(u,v)=(0,1)$ and $(u,v)=(\infty,0)$, and there are two curves on ${\Bbb P}^1\times{\Bbb P}^1$ that contract to a point under its action: $\{v=1\}$ and $\{v=0\}$. Under the action of the mapping, the first of these curves gives rise to the confined singularity pattern of length six described in section 2,
$$\{v=1\}=\pmatrix{f \cr 1}\to\pmatrix{1\cr 0}\to\pmatrix{0\cr\infty}\to\pmatrix{\infty\cr\infty}\to\pmatrix{\infty\cr 0}\to\pmatrix{0\cr 1}\to\{u=1\},\eqdef\antonv$$
which can be read off from the $v$-coordinates in the bottom row of the vectors in the middle of the pattern. The second curve is part of a cyclic pattern of length seven
$$\{v=0\}\to\pmatrix{0\cr\infty}\to\pmatrix{\infty\cr\infty}\to\pmatrix{\infty\cr 0}\to\{u=0\}\to\{v=\infty\}\to\{u=\infty\}\to\{v=0\}\to\cdots,\eqdef\antonvi$$
which, after blow-up, gives rise to a cycle of curves on a rational elliptic surface $X$,
$$\cdots\to\{v=0\}\to C_{0,\infty}\to C_{\infty,\infty}\to C_{\infty, 0}\to\{u=0\}\to\{v=\infty\}\to\{u=\infty\}\to\{v=0\}\to\cdots\eqdef\antonvii$$
(where $C_{0,\infty}, C_{\infty,\infty}$ and $C_{\infty, 0}$ are curves on $X$ obtained from the blow-ups at the points indicated, as shown in Figure 1).  This cycle corresponds of course to the length-seven cyclic pattern we discovered by singularity analysis, for mapping (\zhep). More importantly however, it is clear that the union of all seven curves in this cycle, in fact, constitutes an (invariant) reducible singular fibre on $X$ for the value $K=\infty$. 

\begin{figure}
\centering
\includegraphics[width=7.5cm]{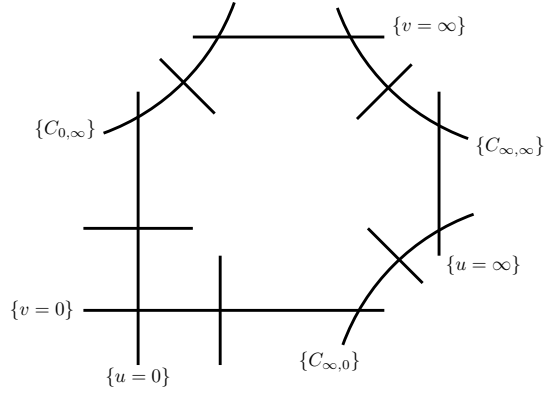}
\caption{The surface obtained after 8 blow-ups for mapping (\zhep)}
\end{figure}

The intersection diagram for this reducible fibre--obtained by associating a vertex with each component of the fibre and with each intersection of the components an edge between the corresponding vertices--is easily seen to form a heptagon, i.e. the Dynkin diagram for A$_6^{(1)}$. The mapping induced on $X$ acts as a rotation of this diagram, as shown in Figure 2.

\begin{figure}
\centering
\centerline{\includegraphics[width=7cm]{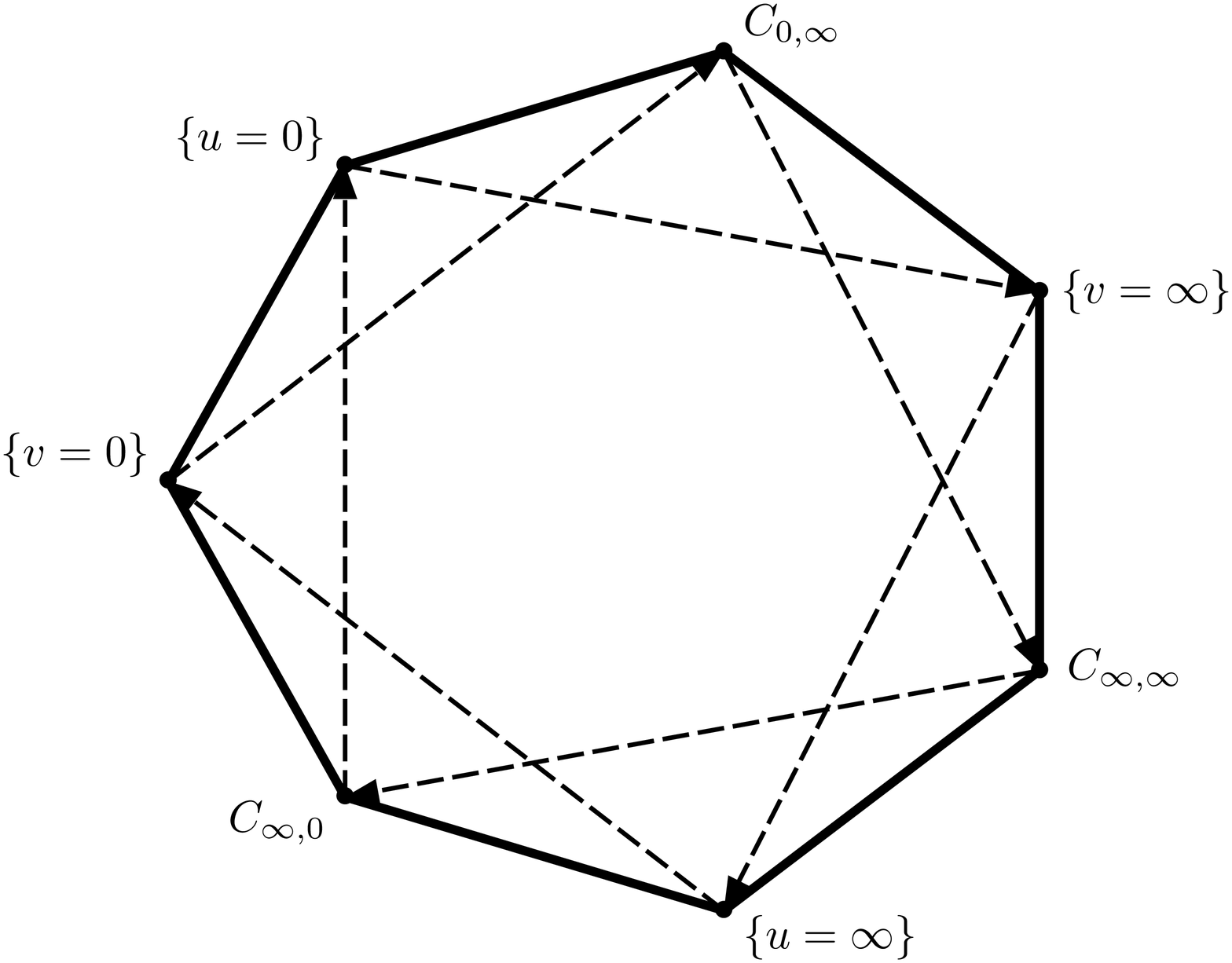}}
\caption{The intersection diagram for the surface shown in Figure 1}
\end{figure}

Oguiso and Shioda in [\os] list only two surfaces that contain such a singular fibre: one where it is the sole reducible fibre to exist on the surface (number 25 in their list) and  one where it is accompanied by an A$_1^{(1)}$ fibre (number 47). It is easy to check directly on the invariant (\znov) itself that there is (exactly) one other reducible fibre: for $K=0$ we find that the invariant decomposes into two curves (on ${\Bbb P}^1\times{\Bbb P}^1$)
$$C_1 :\quad u v = -A\qquad\quad{\rm and}\quad\qquad C_2 :\quad u+v = 1,\eqdef\antonviii$$
which are interchanged by the mapping: the image under the mapping of $C_1$ is $C_2$ and vice versa. After blow-up, these two curves are the components of a singular fibre of $A_1^{(1)}$ type for the surface $X$, but we need to distinguish two cases. When $A\neq-1/4$, $C_1$ and $C_2$ intersect transversally at two distinct points, whereas when $A=-1/4$ they are tangential at a single point. The former type of fibre is usually denoted as A$_1^{(1)}$ and the latter, its so-called additive reduction, as A$_1^{(1)*}$.

In order to better understand the deautonomisation of the mapping (\zhep) that led to (\zoct), i.e. to an equation associated with the affine Weyl group (A$_1$+A$_1)^{(1)}$, it is important to notice first that the set of coordinates (in fact, the homography (\zsex)) we have chosen is such that the curves and singular points in the cyclic pattern (\antonvi) are all independent of the parameter in the equation. Hence, after blow-up, the resulting A$_6^{(1)}$ fibre $\{v=0\} \cup C_{0,\infty}\cup C_{\infty,\infty}\cup C_{\infty, 0}\cup \{u=0\}\cup \{v=\infty\}\cup \{u=\infty\}$ will be, in this sense, completely `fixed' on $X$. (As we shall see in the following this situation might be too restrictive; it suffices to require that the intersections of the components of the fibre are independent of the parameters). This is important since, when deautonomising the mapping, we wish to keep the intersection pattern of the components of the fibre independent of $n$. Now, when deautonomising, the base points (for the blow-ups that lead to $X$) which depend on the parameters of the mapping will also be $n$ dependent, in which case, strictly speaking, we do not obtain a single algebraic surface $X$, but rather a family $\{X_n\}$ of such surfaces. A deautonomisation of an integrable mapping is only meaningful [\refdef\mase] when the map induced between the Picard lattices ${\rm Pic}X_n$ and ${\rm Pic}X_{n+1}$ by its deautonomisation is an isomorphism and, moreover, when this induced action actually coincides with that induced by the original map on ${\rm Pic}X$. In [\cdt] it is explained how, under these circumstances, any given invariant fibre for a QRT-map (smooth or singular) can be  used as an anti-canonical divisor that will define the generalized Halphen surface (in the sense of [\sakai]) that is associated with the discrete Painlev\'e equation obtained from that deautonomisation. 

In a nutshell, in the present case, what happens is that since the isomorphism between ${\rm Pic}X_n$ and ${\rm Pic}X_{n+1}$ is the same as that induced by the original autonomous map (\zhep) on ${\rm Pic}X$, the unitary action on the part of ${\rm Pic}X$ that contains the components of the A$_6^{(1)}$ fibre -- embodied by the cyclic pattern (\antonvii) -- is still present in the nonautonomous setting. Choosing this particular singular fibre as the basic geometric structure to be preserved under deautonomisation, forces the base points for the blow-ups to follow the same permutation pattern as the components of the singular fibre to which they belong. In general, the movement of the base points under the action of the map imposes conditions on the parameters in the map that appear in the coordinates of the base points. However, it turns out that these conditions--equivalent to the confinement conditions--only pick up information about the movement of curves relative to this global periodic movement of the base points. Hence, in this particular deautonomisation of  (\zhep), the period 7 parameter dependence that was present in the original map (\zunu), subject to (\parsi), can no longer be present, but the period 2 dependence in the parameters that is associated with the movement of the components $C_1$ and $C_2$ of the A$_1^{(1)}$ fibre (\antonviii)  does persist in (\zoct).

Let us turn our attention now to the deautonomisations associated with the A$_1^{(1)}$ and A$_1^{(1)*}$ singular fibres. Under the homography (\dtre), $u=(1-z U)/((z+1/z)(z-U))$ and a similar transformation for $v$ with $A=-1/(z+1/z)^2$  $(z^2\neq 1)$, the curves $C_1$ and $C_2$ given by (\antonviii) are transformed into
$$\tilde C_1 :\quad U V = 1\qquad\quad{\rm and}\quad\qquad \tilde C_2 :\quad U V = z^2,\eqdef\antonix$$
which intersect transversally at $(U,V) = (0,\infty)$ and $(\infty,0)$. Identifying $y_{n-1}$ with $U$ and $y_n$ with $V$ in the invariant (\dcua) for the mapping (\dqui), it is obvious that these two curves are the two components of the A$_1^{(1)}$ type singular fibre associated with this particular canonical form of the invariant for the value $K=\infty$. (It should be noted that the invariant ($\dcua$) is in fact proportional to the reciprocal of the original invariant (\znov), which explains why this singular fibre is obtained for the value $K=\infty$ and not at $K=0$ as was originally the case).
The generalized Halphen surface obtained by regularising the deautonomisation (\dhep) of mapping (\dqui), by blow-up, is therefore of type  A$_1^{(1)}$, which is perfectly consistent with the fact that (\dhep) is a multiplicative Painlev\'e equation associated with the E$_7^{(1)}$ Weyl group.

Similarly, the homography (\zdec), $u=(U+z)/(2(U-z)), v=(V+z)/(2(V-z))$ ($z\neq0$) for $A=-1/4$, transforms the curves $C_1$ and $C_2$ to two curves which resemble the ultra-discrete limits of (\antonix):
$$C_1^* :\quad U + V = 0\qquad\quad{\rm and}\quad\qquad C_2^* :\quad U  + V = 2 z.\eqdef\antonx$$
The curves $C_1^*$ and $C_2^*$ intersect tangentially at the point $(\infty,\infty)$ and are, obviously, the two components of the singular fibre associated with the invariant (\dunu) for the value $K=\infty$. Hence, the deautonomisation (\dsex) of mapping (\dduo) is associated with a generalized Halphen surface of type A$_1^{(1)*}$, as indicated by the fact that it is an additive discrete Painlev\'e equation with E$_7^{(1)}$ Weyl group symmetry.

 As expected, in both cases the period 2 parameter dependence that was originally present in (\parsi) has now disappeared, but the period 7 behaviour in the parameters related to the A$_{6}^{(1)}$ fibre has resurfaced.
Note also that in (\antonix) as well as in (\antonx), one of the components of the singular fibre now does depend on the parameter $z$ of the mapping, but that the intersection points do not. Similarly, in the new coordinates we chose for these two mappings, the components of the  A$_{6}^{(1)}$ singular fibre will now depend on the parameter $z$, which makes that the values of the parameters will not match up perfectly and no cyclic pattern of length 7 will show up in the singularity analysis of the nonautonomous mappings. 

As we saw in section 2, these are however not the only possible deautonomisations for the mapping (\zhep). A generic, smooth (so-called A$_0^{(1)}$ type) fibre for the elliptic fibration obtained from the invariant will result in an elliptic discrete Painlev\'e equation such as (\ftre). Furthermore, closer inspection of the detailed lists in [\pers] (or [\mir]) shows that there are in fact three types of rational elliptic surfaces that contain an A$_6^{(1)}$, A$_1^{(1)}$ configuration (allowing for possible additive cases). The first surface, with a configuration of A$_6^{(1)}$ A$_1^{(1)}$ singular fibres and three nodal curves (denoted as 3A$_0^{(1)*}$), corresponds in our case to a generic choice of the parameter $A$ in the QRT-mapping (\zhep), i.e. $A\neq -1/4$ or $27/4$. For each such value of $A$ for the homography (\doct) we find indeed three different values of $\kappa$ (due to the symmetry of the latter under $z\to -z$ and $z\to1/z$), that is three different singular fibres, but the form of the mapping does not change.
For a given generic $A$, by deautonomising (\zhep), we have one $q$-Painlev\'e equation in E$_8^{(1)}$, one in E$_7^{(1)}$ and one in (A$_1$+A$_1)^{(1)}$ (equations (\ddec), (\dhep) and (\zhep) with (\zoct) respectively, obtained in section 2), which are indeed associated to A$_0^{(1)*}$, A$_1^{(1)}$ and  A$_6^{(1)}$ type generalized Halphen surfaces in Sakai's classification. 

A second surface has a configuration A$_6^{(1)}$ A$_1^{(1)*}$ 2A$_0^{(1)*}$, i.e. with one additive type A$_1^{(1)}$ type fibre, which in our case corresponds to the choice $A=-1/4$ in (\zhep). For this value of $A$ we have now two different values of $\kappa$ (again taking into account the invariances of $\kappa$) which are compatible with the homography (\doct).
This choice of the value of $A$ does not affect the equations with E$_8^{(1)}$ or (A$_1$+A$_1)^{(1)}$ symmetry, but the equation with E$_7^{(1)}$ symmetry, obtained after deautonomisation, is now of additive type (equation (\dduo)). 

The final type of surface has the configuration A$_6^{(1)}$ A$_1^{(1)}$ A$_0^{(1)**}$ A$_0^{(1)*}$, where  A$_0^{(1)**}$ stands for a cuspal curve (additive A$_0^{(1)}$ type). This is our case $A=27/4$, for which we find now one value of $\kappa$ that still gives rise (after deautonomisation) to a multiplicative equation of E$_8^{(1)}$. An additive Painlev\'e equation with E$_8^{(1)}$ symmetry does of course exist: it was our starting equation (\zunu), subject to (\parsi).

Finally, it is interesting to point out that as the mapping (\vduo) was obtained by doubling the step in (\zhep), it possesses two invariant curves, 
$$a u v -u - a = 0\qquad{\rm and}\qquad a u v- v - a = 0,\eqdef\antonxi$$
where $u=X_{n-2}, v=X_n$. These are obtained from the factors in the numerator of the invariant  (\vunu), i.e. from the fibre $K=0$,  and are obviously relics of the curves $C_1$ and $C_2$ that were permuted under the original mapping. As was pointed out in section 2, mapping (\vduo) still possesses a cyclic pattern of length 7 but unsurprisingly, its deautonomisation has lost the period 2 dependence in the parameters. 

\bigskip
4. {\scap The symmetric equation with periods 2, 3 \& 5}
\medskip

We shall now investigate the degeneration cascade of the symmetric equation (\zunu) with the parameter dependence (\parsii), along the lines set out in the previous sections. In order to analyse the singularity properties of this map it is in fact sufficient to consider the trihomographic expression 
$${x_{n+1}-(5t_n-\alpha)^2\over x_{n+1}-(t_n+\alpha)^2}{x_{n-1}-(5t_n+\alpha)^2\over x_{n-1}-(t_n-\alpha)^2}{x_{n}-t_n^2\over x_{n}-49t_n^2}=1,\eqdef\vsex$$
with only secular behaviour in the parameters.

Two singularity patterns exist, one of length eight, $\{x_{n-3}=49t_{n-3}^2, x_{n-2}=(5t_{n-3}-\alpha)^2,x_{n-1}=(3t_{n-3}-4\alpha)^2,x_{n}=(t_{n-3}-9\alpha)^2,x_{n+1}=(t_{n+4}+9\alpha)^2, x_{n+2}=(3t_{n+4}+4\alpha)^2,x_{n+3}=(5t_{n+4}+\alpha)^2 ,x_{n+4}=49t_{n+4}^2\}$ and one of length two, $\{x_{n}=t_{n}^2,x_{n+1}=t_{n+1}^2\}$.

Going to the autonomous limit, taking $\alpha=0$ and scaling $\beta$ to 1, we find the mapping
$${x_{n+1}-25\over x_{n+1}-1}{x_{n-1}-25\over x_{n-1}-1}{x_{n}-1\over x_{n}-49}=1.\eqdef\vhep$$
Introducing a new dependent variable $X$ by
$${1\over2}X_n={x_n-25\over x_n-1},\eqdef\voct$$
we find the mapping
$$X_{n+1}X_{n-1}=A(X_n-1),\eqdef\vnov$$
where $A=4$.  The singularity pattern starting with $x=49$ corresponds to a pattern starting with $X=1$ found to be $\{1,0,-A,\infty, \infty,-A,0,1\}$. The length-two pattern, starting with $x=1$, corresponds to one starting with $X=\infty$. More precisely, starting from a generic (finite) value $f$ for $X_{n-1}$ and with $X_n=\infty$, we find the cyclic pattern $\{f,\infty,\infty,f',0,f'',\infty,\infty,\dots\}$ (where $f,f',f'',\dots$ are finite values depending on $X_{n-1}$), consisting of a repetition of the length-five pattern $\{f,\infty,\infty,f',0\}$. Again, the length-two confined pattern expected from that of (\vsex), i.e. $\{\infty,\infty\}$ in the variable $X_n$, is embedded in the cyclic one, bracketed by two finite values. However, it is worth pointing out that if, with hindsight, one would have continued the iteration for the mapping (\vsex) starting from $x_{n-1}=w, x_n=t_{n}^2$ for a few steps more, i.e. beyond the point where the singularity actually confines, one would have found the succession of values
$$x_{n-1}=w,~ x_{n}=t_{n}^2, ~x_{n+1}=t_{n+1}^2, ~x_{n+2}= g(w), ~x_{n+3} = ~(5 t_{n+2}-\alpha)^2, x_{n+4} =~g'(w), ~\dots,\eqdef\remxx$$
(where $g(w)$ and $g'(w)$ are complicated functions of the initial value $w$) which corresponds to the succession $f,\infty,\infty,f',0,f'', \dots$ in the autonomous case (\vnov). This shows that the nonautonomous mapping (\vsex) in fact picks up the slightly longer subpattern $\{f,\infty,\infty,f',0,f''\}$ from the cyclic pattern than is apparent from a simple check of the singularity confinement property.

Assuming now $A$ to be a function of $n$ we can try to deautonomise (\vnov). The singularity confinement constraint yields
$$A_{n+3}A_{n-2}=A_{n+1}A_n,\eqdef\tdec$$
which can be integrated to $A_n=\alpha n+\beta +\phi_2(n)+\phi_3(n)$. The corresponding discrete Painlev\'e equation is thus associated to the affine Weyl group A$_4^{(1)}$ [\mulkmt] . Although the period 5 behaviour in the parameters has disappeared compared to those of (\zunu) with (\parsii), the singularity analysis shows that the length-5 cyclic pattern persists even when (\vnov) is deautonomised. 

The mapping (\vnov), being of QRT-type, has an invariant of the form
$$K={(X_nX_{n-1}+A(X_n+X_{n-1})-A)(X_n+X_{n-1}+A)\over X_nX_{n-1}},\eqdef\tunu$$
from which it is obvious that the cycle $\{v=\infty\} \to (\infty,\infty) \to \{u=\infty\} \to \{v=0\} \to \{u=0\} \to  \{v=\infty\} \to \cdots$ (where $u=X_{n-1}$ and $v=X_n$)  is nothing but the footprint on ${\Bbb P}^1\times{\Bbb P}^1$ of an A$_4^{(1)}$ type singular fibre associated with (\tunu) at $K=\infty$. As explained in section 3, it is this fibre which underlies the geometric structure of the deautonomisation, which is therefore associated with a surface of A$_4^{(1)}$ type, resulting in a discrete Painlev\'e equation with A$_4^{(1)}$ Weyl group symmetry. Here again, as the curves in this singular fibre do not depend on the parameter $A$ of the mapping, the cyclic pattern resulting from the invariance of the fibre still persists in the nonautonomous equation.

As in section 2, we can now use the invariant (\tunu) as a starting point to look for the possible canonical forms that can be obtained from (\vnov), through a homographic change of the dependent variable and a shift of the value of the invariant by adding a constant $\kappa$. First we must discard the trivial case $A=-1$ which corresponds to a mapping with a periodic, period-5, solution. Then, in total, six canonical cases are left. First, for $A\neq-1/2$, introducing the homography
$$X_n=-{Ay_n+A+1\over y_n-1},\eqdef\tduo$$
we obtain the invariant (with $\kappa=-1-A$)
$$K={(y_n-1)(y_{n-1}-1)(Ay_n+A+1)(Ay_{n-1}+A+1)\over y_ny_{n-1}-1},\eqdef\ttre$$
corresponding to the mapping
$$(y_ny_{n+1}-1)(y_ny_{n-1}-1)=(1-y_n)(1+y_n(1+1/A)).\eqdef\tcua$$
Rewriting the right-hand side of (\tcua) as $(1-y_n)(1-z_ny_n)$ we find that, for the singularities of the mapping to confine, $z_n$  must satisfy the constraint 
$$z_{n+5}z_{n-2}=z_{n+3}z_n,\eqdef\tqui$$
the integration of which leads to $\log z_n=\alpha n+\beta+\phi_2(n)+\phi_5(n)$ [\mulkmt]. The corresponding $q$-discrete Painlev\'e equation is associated to the affine Weyl group E$_6^{(1)}$ (as the period 3 dependence that was originally present in (\parsii) has now disappeared). It is easily checked on the invariant (\ttre) that there is a singular fibre for $K=\infty$ that decomposes into three curves (on ${\Bbb P}^1\times{\Bbb P}^1$) which form the 3-cycle $~\cdots \to \{u=\infty\} \to \{u v=1\} \to \{v=\infty\} \to \cdots~$
(where $u=y_{n-1}$ and $v=y_n$). Since the intersection of these curves in non-degenerate, this singular fibre is of type A$_2^{(1)}$, which indeed gives the correct surface in Sakai's classification. Note also that this 3-cycle does not give rise to a cyclic singularity pattern for mapping (\tcua), since it does not contain any singularities (none of the curves in it contract to a point under the action of the mapping). 

The situation changes when $A=-1/2$. In this case we use the homographic transformation 
$$X_n={y_n+z\over 2y_n},\eqdef\tsex$$
which, with $\kappa=-1/2$, yields an invariant of the form
 $$K={y_ny_{n-1}(y_n+z)(y_{n-1}+z)\over y_n+y_{n-1}}.\eqdef\thep$$
The corresponding mapping is now of additive type:
$$(y_n+y_{n+1})(y_n+y_{n-1})=y_n(y_n-z).\eqdef\toct$$
Deautonomising (\toct) by applying the confinement criterion we find for $z$ the constraint
$$z_{n+5}+z_{n-2}=z_{n+3}+z_n,\eqdef\tnov$$
and $z_n=\alpha n+\beta+\phi_2(n)+\phi_5(n)$ [\refdef\addkmt]. Hence, the discrete Painlev\'e equation we obtain is again associated to the affine Weyl group E$_6^{(1)}$, but is now of additive type.  Indeed, the singular fibre obtained from (\thep) for $K=\infty$, is this time given by three lines that intersect at a single point, $(\infty,\infty)\in{\Bbb P}^1\times{\Bbb P}^1$, and that form the 3-cycle $~\cdots \to \{u=\infty\} \to \{u+v=0\} \to \{v=\infty\} \to \cdots~$ and this fibre is therefore of A$_2^{(1)*}$ type, i.e. of additive type. We would like to point out that both discrete Painlev\'e equations associated to E$_6^{(1)}$ have been first identified in [\refdef\capel] and then in [\mulkmt] and [\addkmt] respectively.

The next homographic transformation is
$$X_n={ay_n-b\over y_n-1},\eqdef\tdec$$
where the constants $a,b ~(a\neq b)$ are related by the constraint $a^2+ab+b^2-a-b=0$. The corresponding invariant, obtained for $A=-(a+b)$ and $\kappa=0$, is
$$K={(ay_n-b)(ay_{n-1}-b)(y_n-1)(y_{n-1}-1)\over(by_ny_{n-1}-a)(y_ny_{n-1}-1)},\eqdef\qunu$$
leading to the mapping
$$\left(y_ny_{n+1}-a/b\over y_ny_{n+1}-1\right)\left(y_ny_{n-1}-a/b\over y_ny_{n-1}-1\right)={y_n-a^2/b^2\over y_n-1}.\eqdef\qduo$$
An additive mapping can be similarly obtained by taking 
$$X_n={2\over3}{y_n+z\over y_n},\eqdef\qcua$$
$(z\neq0)$ with $A=-4/3$ and $\kappa=0$, which leads to the invariant
$$K={y_ny_{n-1}(y_n+z)(y_{n-1}+z)\over(y_n+y_{n-1}-z)(y_n+y_{n-1})},\eqdef\qqui$$
and the mapping 
$$\left(y_n+y_{n+1}-z\over y_n+y_{n+1}\right)\left(y_n+y_{n-1}-z\over y_n+y_{n-1}\right)={y_n-2z\over y_n}.\eqdef\qsex$$
Note that a different choice for (\qcua), $X_n=(2y_n-3z)/(3y_n-3z/2)$,  leads to a mapping dual to (\qcua) where the right-hand side is $(y_n-z/2)/(y_n+3z/2)$. 

The deautonomisation of (\qduo) and (\qsex) has been performed in [\refdef\sequel] where we have shown that the resulting discrete Painlev\'e equations are related to the affine Weyl group E$_7^{(1)}$, the periodicities of their coefficients being 3 and 5. The period 2 behaviour that (compared to the original mapping) has disappeared from the parameters, is related to the singular fibre $ \{uv=1\}\cup\{b\;u v = a\}$ for $K=\infty$ that is apparent on the canonical form of the invariant (\qunu) (the components of which are obviously interchanged by the mapping (\qduo)) or to the singular fibre $\{u+v=0\} \cup \{u+v=z\}$ for the invariant (\qqui) of (\qsex), again for $K=\infty$. Note that, as before, although the components of these fibres depend on the parameter $z$, their intersections do not. The singular fibre for equation (\qduo) intersects in two different points and is therefore of type A$_1^{(1)}$, whereas that for (\qsex) is of type A$_1^{(1)*}$, as was to be expected.

Further choices for the homographic transformations do exist. Obviously when $A=4$ the homography (\voct) takes us back to the additive E$_8^{(1)}$ mapping we started with. A multiplicative mapping can be obtained by introducing the homography
$$X_n={z^2+1/z^2\over z^2-1+1/z^2}{y_n+z^5+1/z^5\over y_n+z+1/z},\eqdef\qhep$$
and the values $A=(z^2+1/z^2)^2/(z^2-1+1/z^2)$, $\kappa=-(z^2+1/z^2)(z+1/z)^4/(z^2-1+1/z^2)$. The corresponding invariant is
$$K={(y_n+z+1/z)(y_{n-1}+z+1/z)(y_n+z^5+1/z^5)(y_{n-1}+z^5+1/z^5)\over y_n^2+y_{n-1}^2-(z^2+1/z^2)y_ny_{n-1}+(z^2-1/z^2)^2},\eqdef\qoct$$
and the mapping has the form
$${(y_n-z^2y_{n+1})(y_n-z^2y_{n-1})-(z^4-1)^2\over (z^2y_n-y_{n+1})(z^2y_n-y_{n-1})-(z^4-1)^2/z^4}={y_n+z^4(z^3+1/z^3)\over z^4y_n+z^3+1/z^3}.\eqdef\qnov$$
By deautonomising (as we did in [\multi]) we obtain a multiplicative discrete Painlev\'e equation with dependence (\parsii) in the coefficients, associated to the affine Weyl group E$_8^{(1)}$. 

At this point it is worth summarising the results we obtained so far. In total, discarding the case $A=-1$ which leads to a periodic mapping, we have found four different cases for the value of the parameter $A$ in mapping (\vnov). When $A$ is generic, i.e. $A\neq 4, -1/2$ or $-4/3$, we found that (\vnov) can be deautonomised to $q$-Painlev\'e equations with E$_8^{(1)}$, E$_7^{(1)}$, E$_6^{(1)}$ and A$_4^{(1)}$ symmetries (the deautonomisations of equations (\qnov), (\qduo), (\tcua)  and (\vnov) respectively). For such generic values of $A$, the QRT-map (\vnov) is therefore associated with a rational elliptic surface with an A$_4^{(1)}$ A$_2^{(1)}$ A$_1^{(1)}$ 2A$_0^{(1)*}$ configuration of singular fibres. When $A=4$, the equation with E$_8^{(1)}$ symmetry is of additive type (equation (\zunu) with (\parsii)) and the corresponding configuration of singular fibres on the elliptic surface for (\vnov) becomes A$_4^{(1)}$ A$_2^{(1)}$ A$_1^{(1)}$ A$_0^{(1)**}$. On the other hand, when $A=-4/3$, the multiplicative equation with E$_8^{(1)}$ symmetry still exists but the E$_7^{(1)}$ one is now additive (equation (\qsex)), which corresponds to a A$_4^{(1)}$ A$_2^{(1)}$ A$_1^{(1)*}$ A$_0^{(1)*}$ configuration on the elliptic surface. The final configuration is A$_4^{(1)}$ A$_2^{(1)*}$ A$_1^{(1)}$ A$_0^{(1)*}$, obtained for $A=-1/2$, for which the equation with E$_6^{(1)}$ symmetry becomes additive (equation (\toct)). These four types of configurations exhaust all possibilities for such a surface, as can be verified in Persson's list [\pers] and as has been explained in great detail in the monograph of Duistermaat, chapter 11.4 [\refdef\duist].

Obviously an elliptic discrete Painlev\'e equations can also be constructed, following the same procedure we presented in section 2. We shall not go here into these details. The calculations are elementary and the interested readers can repeat them for themselves.

Just as in the case of the periods 2 and 7 mapping we can consider double-step evolutions. We start from (\vnov), solve for $X_n$ and, after a downshift of the indices, we obtain the invariant
$$K={(X_nX_{n-2}-aX_n-1+a^2)(X_nX_{n-2}-aX_{n-2}-1+a^2)\over X_nX_{n-2}-1},\eqdef\qdec$$
where we have taken $A=-a^2$ and have rescaled $X$ so as to absorb a factor of $a$. The resulting mapping is
$$(X_nX_{n+2}-1)(X_nX_{n-2}-1)= a^2(1-aX_n).\eqdef\punu$$
Its singularity patterns are $\{1/a,a,\infty,0\}$, its mirror image $\{0,\infty,a,1/a\}$ and the cyclic pattern of length five $\{f,\infty,0, \infty, f',1/f', \infty, 0, \cdots\}$.

 In order to deautonomise (\punu) we rewrite it as $(X_nX_{n+2}-1)(X_nX_{n-2}-1)=b_n(1-a_nX_n)$. Using the singularity confinement criterion we find that $b_n=da_{n+1}a_{n-1}$, where $d$ is a constant, while $a_n$ obeys the relation $a_{n+4}a_n=a_{n+3}a_{n+1}$ the solution of which is $\log a_n=\alpha n+\beta+\phi_3(n)$. This result was first obtained in [\mulkmt], equation 124. The period 2 that was present in the parameter dependence in the original mapping (\zunu), subject to (\parsii), has disappeared here because of the doubling of the step. Similarly the period 5 has disappeared because this deautonomisation involves the A$_4^{(1)}$ singular fibre that underlies the cyclic pattern we found for (\punu). 

Introducing the invariant associated to the autonomous limit of (\punu) keeping the constant $d$ and putting $h=a^2d$ we find
$$K={X_n^2X_{n-2}^2-a X_nX_{n-2}(X_n+X_{n-2})+a(1-h)(X_n+X_{n-2})+a^2h+h-1\over X_nX_{n-2}-1}.\eqdef\aduo$$
Using this invariant we can now derive the double-step equivalent of (\tcua) and (\toct). Starting from the additive equation we introduce the homography
$$X_n=a{y_n-z\over y_n+z},\eqdef\bunu$$
and, under the constraint $h=1-a^2$, we obtain, for $\kappa=0$, the invariant 
$$K={(y_ny_{n-2}-mz(y_n+y_{n-2})+z^2)(y_n+z)(y_{n-2}+z)\over y_n+y_{n-2}},\eqdef\bduo$$
where $m=(a^2+1)/(a^2-1)$. The corresponding mapping is
$$(y_n+y_{n+2})(y_n+y_{n-2})={(y_n+z)(y_n-z)^2\over y_n-mz}.\eqdef\btre$$
The deautonomisation of this mapping was presented in [\addkmt], equation (35), which has coefficients with period five as well as one free parameter. 

Next we turn to the multiplicative equation. We introduce the homography
$$X_n=a{y_n+1/c\over y_n+c},\eqdef\bcua$$
and provided we set $h= 1-a^2/c^2$ we find the invariant (with $\kappa=a^2 (1-1/c^2)$)
$$K={(c^2 (a^2-1) y_n y_{n-2} + c (a^2-c^2) (y_n + y_{n-2}) + a^2 -c^4)(y_n+c)(y_{n-2}+c)\over y_ny_{n-2}-1}.\eqdef\bqui$$
The multiplicative mapping is now
$$(y_ny_{n+2}-1)(y_ny_{n-2}-1)={(1+y_n/c)(1+cy_n)^2\over 1+c \mu y_n},\eqdef\bsex$$
where we have introduced the auxiliary quantity $\mu=(a^2-1)/(a^2-c^2)$. 
We have obtained the deautonomisation of this mapping in [\mulkmt], equation (51), which again has coefficients with period five as well as one free parameter. 

In order to obtain double step equation starting for the E$_7^{(1)}$-associated ones we start from the multiplicative equation (\punu). (Using the invariant (\aduo) would not have led to a different result since a constraint for the obtention of the E$_7^{(1)}$-associated equations is $d=1$). Solving (\qduo) for $y_n$ we find from (\qunu) the invariant
$$K={(y_{n-2}-1)(y_n-1)\big((z+1)y_{n-2}y_n-(1+z^2)(y_{n-2}+y_n)+z\big)\over(zy_n-y_{n-2})(zy_{n-2}-y_n)},\eqdef\pduo$$
where we have taken $a/b=z$. The corresponding mapping is
$$\left({zy_{n+2}-y_n\over y_{n+2}-zy_n}\right)\left({zy_{n-2}-y_n\over y_{n-2}-zy_n}\right)={y_n^2-(1+z^2)y_n+z^2\over (zy_n-1)^2},\eqdef\ptre$$
again given in (VI$'$) form. Its deautonomisation, in the canonical form (VI), was given in [\sequel]. As expected, the same procedure applied to the additive-type mapping (\qsex) leads to a (V$'$) form
$$\left({y_{n+2}-y_n+z\over y_{n+2}-y_n-z}\right)\left({y_{n-2}-y_n+z\over y_{n-2}-y_n-z}\right)={y_n(y_n-2z)\over (y_n+z)^2}.\eqdef\pcua$$
In order to obtain the double-step equations related to E$_8^{(1)}$ group we start from the invariant (\aduo), obtain the double step mapping and deautonomise it. We find that equation (\vsex), which corresponds to case 3.1 in [\ancil], is related, once we consider the double step evolution, to case 4.3.4 of the same paper. By performing a similar analysis we can obtain a multiplicative double step equation and, in fact, also an elliptic one. 

It turns out that for the present system it is also possible to consider a triple step evolution. Starting from the invariant (\tunu) we obtain 
$$K={(AX_nX_{n-3}-A-1)(AX_nX_{n-3}-X_n-X_{n-3}-A-2)\over X_nX_{n-3}-1},\eqdef\triena$$ 
where we have absorbed one factor of $A$ in $X$ so as to bring the denominator to a canonical form. The corresponding mapping is
$$(X_nX_{n-3}-1)(X_nX_{n+3}-1)={1\over A^2}{(X_n+1)^2\over X_n-1/A}.\eqdef\tridyo$$
The deautonomisation of this mapping leads to a right-hand side of the form $az_nz_{n+1}(X_n^2-(b+1/b)X_n+1)/(X_n-z_n)$  where $\log z_n=\alpha n+\beta+\gamma(-1)^n$ and $a,b$ are constant. The form of its left-hand side notwithstanding, this mapping is indeed in A$_4^{(1)}$, a result obtained in [\mulkmt]. The autonomous form we should therefore work with is
$$(X_nX_{n-3}-1)(X_nX_{n+3}-1)=a{(X_n-b)(X_n-1/b)\over X-c},\eqdef\tridyo$$
corresponding to the invariant
$$K={X_n^2X_{n-3}^2-cX_nX_{n-3}(X_n+X_{n-3})+(a+c)(X_n+X_{n-3})-a(b+1/b)-1\over X_nX_{n-3}-1}.\eqdef\tritri$$
By implementing the proper homographic transformation we can now obtain from (\tritri) a mapping which, when deautonomised, leads to an equation associated to the affine Weyl group E$_7^{(1)}$. We shall not go into all the details but just give the corresponding multiplicative discrete Painlev\'e equation. It has the form
$$\displaylines{\left({y_ny_{n+3}- z_{n-1} z_{n} z_{n+1} z_{n+2} z_{n+3} z_{n+4}\over y_ny_{n+3}-1}\right)\left({y_ny_{n-3}-  z_{n-4} z_{n-3} z_{n-2} z_{n-1} z_{n} z_{n+1}\over y_ny_{n-3}-1}\right)
\hfill\cr\hfill= {(y_n- z_{n} z_{n+1} z_{n+4})(y_n- z_{n-4} z_{n-1} z_{n})(y_n- z_{n-2} z_{n-1} z_{n} z_{n+1} z_{n+2})\over (y_n-z_n ) (y_n-1/z_{n+3}) (y_n-1/z_{n-3}) }, \quad\eqdisp\trites \cr}$$
where $\log z_n=\alpha n+\beta+\phi_3(n)+\phi_5(n)$. Note that since the evolution takes place with a triple step the $\phi_3(n)$ term could have been neglected in $z_n$, to be replaced by two constants in the appropriate places in the equation.  
Similarly we can find the corresponding equation of additive type
$$\displaylines{\left({y_n+y_{n+3}\!-\!z_{n-1}\!-\!z_{n}\!-\!z_{n+1}\!-\!z_{n+2}\!-\!z_{n+3}\!-\!z_{n+4} \over y_n+y_{n+3}} \right)\left({y_n+y_{n-3}\!-\!z_{n-4}\!-\!z_{n-3}\!-\!z_{n-2}\!-\!z_{n-1}\!-\!z_{n}\!-\!z_{n+1}\over y_n+y_{n-3}}\right)\hfill\cr\hfill
= {(y_n-z_{n}-z_{n+1}-z_{n+4})(y_n-z_{n-4}-z_{n-1}-z_{n})(y_n-z_{n-2}-z_{n-1}-z_{n}-z_{n+1}-z_{n+2})\over (y_n-z_n ) (y_n+z_{n+3}) (y_n+z_{n-3}) }, \quad\eqdisp\tripen \cr}$$
where $z_n=\alpha n+\beta+\phi_3(n)+\phi_5(n)$, and the same remark on $\phi_3(n)$ applies here.
Curiously these equations do not seem to have been derived before. 

Triple-step equations associated to E$_6^{(1)}$ can also be obtained. In this case the extension of the invariant to the form (\tritri) is not mandatory, the form (\triena) being sufficient. We give directly the final result for the multiplicative equation
$$\left({x_{n+3}-z_{n+1}x_n\over x_{n+3}z_{n+2}-x_n}\right)\left({x_{n-3}-z_{n-1}x_n\over x_{n-3}z_{n-2}-x_n}\right)={(x_nz_{n+1}-1)(x_nz_{n-1}-1)\over (x_n-1)(x_n-z_n)},\eqdef\trihex$$
where $\log z_n=\alpha n+\beta+\phi_2(n)+\phi_5(n)$ and the additive one
$$\left({x_{n+3}-z_{n+1}-x_n\over x_{n+3}-z_{n+2}-x_n}\right)\left({x_{n-3}-z_{n-1}-x_n\over x_{n-3}-z_{n-2}-x_n}\right) ={(x_n+z_{n+1})(x_n+z_{n-1})\over x_n(x_n-z_n)},\eqdef\trihep$$
with $z_n=\alpha n+\beta+\phi_2(n)+\phi_5(n)$. Again, these two equations do not seem to have been previously derived. Obtaining triple-step equations associated to E$_8^{(1)}$ is also possible but it turns out that the resulting additive equation is the case 5.2.5 of (\ancil) and thus need not be presented explicitly here. (Obviously, once the result for the additive equation are obtained, transcribing it to the multiplicative and elliptic cases is elementary, as explained in the introduction).

\bigskip
5. {\scap The asymmetric equation with period 8}
\medskip

The discrete Painlev\'e equation (\zduo) with period 8 (as given by (\parsiii)) was first derived in [\maxim]. In [\eight] it was again obtained during the complete classification of trihomographic equations associated to the group E$_8^{(1)}$. It can only be expressed in asymmetric (in the QRT parlance) form:
$${x_{n+1}-4t_n^2\over x_{n+1}}{x_{n}-4t_n^2\over x_{n}}{y_{n}-t_n^2\over y_{n}-9t_n^2}=1\eqdaf\pqui$$
$${y_{n}-(3t_n-2\alpha)^2\over y_{n}-t_n^2}{y_{n-1}-(3t_n-\alpha)^2\over y_{n-1}-(t_n-\alpha)^2}{x_{n}\over x_{n}-(4t_n-2\alpha)^2}=1.\eqno(\pqui b)$$
Four singularity patterns do exist: $\{x_n=0,y_n=t_n^2\}$, $\{y_n=t_n^2,x_{n+1}=0\}$, $\{y_{n-1}=9t_{n-1}^2, x_n=4t_{n-1}^2, y_n=(t_{n-1}-\alpha)^2, x_{n+1}=4\alpha^2, y_{n+1}=(t_{n+2}+\alpha)^2, x_{n+2}=4t_{n+2}^2, y_{n+2}=9t_{n+2}^2\}$ and  $\{x_{n-2}=(4t_{n-2}-2\alpha)^2, y_{n-2}=(3t_{n-2}-2\alpha)^2, x_{n-1}=(2t_{n-2}-2\alpha)^2, y_{n-1}=(t_{n-2}-3\alpha)^2, x_n=16\alpha^2, y_n=(t_{n+2}+2\alpha)^2, x_{n+1}=4t_{n+2}^2, y_{n+1}=(3t_{n+2}-\alpha)^2, x_{n+2}=(4t_{n+2}-2\alpha)^2\}$. Next we autonomise (\qqui) obtaining
$${x_{n+1}-4\over x_{n+1}}{x_{n}-4\over x_{n}}{y_{n}-1\over y_{n}-9}=1,\eqdaf\psex$$
$${y_{n}-9\over y_{n}-1}{y_{n-1}-9\over y_{n-1}-1}{x_{n}\over x_{n}-16}=1.\eqno(\psex b).$$
First let us point out that since $y$ and $x$ appear homographically in (\psex a) and (\psex b) we can eliminate either of them and obtain an equation for the other variable. We find thus 
$${(x_{n+1}-x_n-4)(x_{n-1}-x_n-4)+16x_n\over x_{n+1}-2x_n+x_{n-1}-8}=-{x_n(x_n+20)\over2(x_n+2)}\eqdef\atre$$
and
$${(y_{n+1}-y_n-4)(y_{n-1}-y_n-4)+16y_n\over y_{n+1}-2y_n+y_{n-1}-8}=-{y_n^2+18y_n-27\over 2 y_n}.\eqdef\acua$$
They are the autonomous limits of equations 4.2.3 and 4.3.2 of E$_8^{(1)}$ type derived in  [\ancil]. Their deautonomisation was presented in full detail in that paper.

Going back to (\psex) we can introduce two new variables $X$ and $Y$ as
$$X_n={x_n-4\over x_n}, \quad Y_n={y_n-9\over y_n-1},\eqdef\phep$$
and rewrite (\psex) as 
$$X_nX_{n+1}=Y_n,\qquad Y_nY_{n-1}=AX_n+B,\eqdef\miu$$ 
where $A=4$ and $B=-3$, to obtain the invariant 
$$K={X_nY_n^2+AX_n^2+BX_n+AY_n\over X_nY_n}.\eqdef\eunu$$
Equation (\miu) can be easily deautonomised, leading to $\log A_n=\alpha n+\beta$ and $B$ a constant and an equation in A$_1^{(1)}$.
When we view equation (\miu) as a mapping on ${\Bbb P}^1\times{\Bbb P}^1$, $(u,v)\to(v/u,\;A/u + B/v)$ with $u=X_n$ and $v=Y_n$, it is easily verified that there are only 3 curves that contract to a point and give rise to a singularity: $\{A v + B u =0\} ,~\{u=0\}$ and $\{v=0\}$. The first curve is the only one that gives rise to a genuinely confined singularity pattern
$$\{A v + B u =0\}\to (-{B\over A}, 0)\to (0, \infty) \to (\infty^2, \infty) \to (0,0) \to \{u=-{B\over A}\},\eqdef\uvpatti$$
which corresponds exactly to  the length-nine pattern we found for the mapping (\pqui),
$$\{X_n= -{B\over A}, Y_n= X_{n+1}=0, Y_{n+1}=\infty, X_{n+2}=\infty^2, Y_{n+2}=\infty, X_{n+3}=Y_{n+3}=0, X_{n+4}=-{B\over A}\}.\eqdef\uvpattii$$
(The notation $\infty^2$ is shorthand for the following: had we introduced a small quantity $\epsilon$, assuming that $X_{n+1}=\epsilon$, we would have found that $X_{n+2}$ is of the order of $1/\epsilon^2$). The two other curves turn out to be part of a cyclic singularity pattern of length 8:
$$\cdots\to\{u=0\}\to(\infty,\infty)\to\{v=0\}\to(0,\infty)\to(\infty^2,\infty)\to(0,0)\to\{v=\infty\}\to\{u=\infty\}\to\cdots,\eqdef\uvpattiii$$
which is nothing but the footprint on ${\Bbb P}^1\times{\Bbb P}^1$ of the A$_7^{(1)}$ singular fibre that exists for (\eunu) at $K=\infty$. It is easily verified that the three remaining patterns we found for (\pqui) are contained in this cyclic pattern.

Next we look for canonical forms starting from the invariant (\eunu) by introducing two different homographies for $X$ and $Y$. Obviously the homography (\phep) would bring us back to our starting point. However another possibility exists, leading to a multiplicative mapping. Introducing the homography
$$X_n=a{x_n+z^2+1/z^2\over x_n+2}, \quad Y_n=a^2{y_n+z^3+1/z^3\over y_n+z+1/z},\eqdef\eduo$$
and parametrising $A=a^3(z+1/z)^2$, $B=-a^4(z^2+1+1/z^2)$, where the parameter $a$ satisfies the equation $a^4-Aa-B=0$.
 We then obtain, from (\eunu) with $\kappa=-a^2(z^2+4+1/z^2)$, the invariant 
$$K={(x_n+z^2+1/z^2)(x_n+2)(y_n+z^3+1/z^3)(y_n+z+1/z)\over (x^2+y^2)z^2-xyz(z^2+1)+(z^2-1)^2}.\eqdef\etre$$
From the conservation of $K$ we find the system of equations
$${x_{n+1}+z^2+1/z^2\over x_{n+1}+2}{x_{n}+z^2+1/z^2\over x_{n}+2}{y_n+z+1/z\over y_n+z^3+1/z^3}=1,\eqdaf\ecua$$
$${y_n+z+1/z\over y_n+z^3+1/z^3}{y_{n-1}+z+1/z\over y_{n-1}+z^3+1/z^3}{x_{n}+z^4+1/z^4\over x_{n}+2}=1.\eqno(\ecua b).$$
Since these equations have a trihomographic form we can eliminate either $x$ or $y$ and obtain an equation for a single variable. Eliminating $y$ from (\ecua a) and (\ecua b) we find
$${(x_n-z^2x_{n+1})(x_n-z^2x_{n-1})-(z^4-1)^2\over (z^2x_n-x_{n+1})(z^2x_n-x_{n-1})-(z^4-1)^2/z^4}=z^4{x_n^2+x_n(z^2+1)^2-z^8+2z^6+2z^4+2z^2-1\over x_n^2z^8+x_nz^{8}(z^2+1)^2-z^8+2z^6+2z^4+2z^2-1}.\eqdef\equi$$
Similarly, eliminating $x$ we obtain
$${(y_n-z^2y_{n+1})(y_n-z^2y_{n-1})-(z^4-1)^2\over (z^2y_n-y_{n+1})(z^2y_n-y_{n-1})-(z^4-1)^2/z^4}=z^4{y_n^2+y_n(3z^3+1/z)-z^8+3z^6+3z^2-1\over y_n^2z^8+y_nz^5(z^4+3)-z^8+3z^6+3z^2-1}.\eqdef\esex$$
Equations of the form (\equi) and (\esex) where first derived in [\refdef\mickens] where they have been deautonomised albeit only as far as the secular dependence is concerned. Under a complete deautonomisation we expect these two equations to be the multiplicative analogues of the E$_8^{(1)}$ associated equations 4.2.3 and 4.3.2 of [\ancil].

In Persson's classification [\pers] there are two types of configurations that only have an A$_7^{(1)}$ fibre as a reducible fibre, A$_7^{(1)}$ 4A$_0^{(1)*}$ and  A$_7^{(1)}$ A$_0^{(1)**}$ 2A$_0^{(1)*}$. The first case corresponds to the generic situation where the values of $A$ and $B$ are such that $A^4/B^3\neq-256/27$. When this is true, the equation $a^4-Aa-B=0$ has generically four distinct roots giving rise to four different values of $\kappa$. In this case we have two multiplicative Painlev\'e equations, one of A$_1^{(1)}$ and one of E$_8^{(1)}$ type. The second case corresponds precisely to $A^4/B^3=-256/27$. (Notice that Duistermaat, who has also studied this mapping, in chapter 11.7.1 of [\duist], gives a constraint in the form $\delta^5=-8/27$. Unfortunately this value is wrong and our analysis shows that the correct value is $\delta^5=256/27$). Under this constraint we can still have a meaningful homography, like (\eduo), to a multiplicative equation (in E$_8^{(1)}$ when deautonomised) for values of $z$ that satisfy $27z^8+68z^6+98z^4+68z^2+27=0$, which leads to two distinct values for $\kappa$ (given the invariance of the latter under $z\to-z$ and $z\to1/z$). On the other hand, the scaling freedom which exists at the level of the constraint $A^4/B^3=-256/27$ translates itself to a mere scaling freedom in equation (\miu) an thus, up to a rescaling of $X$ and $Y$, the only homography leading to an additive equation of E$_8^{(1)}$ type is (\phep).

A different approach is also possible, but it leads to the same conclusions. Going back to the mapping (\miu) for $X$ and $Y$ we can eliminate $Y$ and find for $X$ the mapping
$$X_{n+1}X_{n-1}={A\over X_n}+{B\over X_n^2}.\eqdef\poct$$
By rescaling the variable $X$ we can always put $B=-1$ which corresponds to the customary form of (\poct). The singularity patterns of (\poct) are $\{-B/A,0,\infty^2,0,-B/A\}$ and the cyclic one $\{0,\infty^2,0,f,\infty,0,\infty, f'\}$ of length 8, which are easily read off from the patterns (\uvpattii) and (\uvpattiii) respectively. The deautonomisation of (\poct) leads to $\log A_n=\alpha n+\beta$, (with $B=-1$), which means that the corresponding discrete Painlev\'e equation has just one degree of freedom and is associated to A$_1^{(1)}$.

The invariant of (\poct) is 
$$K={X_{n}^2X_{n-1}^2+A(X_{n}+X_{n-1})+B\over X_{n}X_{n-1}}.\eqdef\pnov$$
This can be the starting point for the derivation of other canonical forms to be obtained through homographic transformations of the dependent variable. We shall not go into all the details but no such possibility exists apart from forms associated to the group E$_8^{(1)}$ and in fact we find precisely the result obtained above working with the asymmetric form, i.e. equation (\equi).

Similarly, eliminating $X$ from the mapping for (\eunu) we obtain
$$(Y_nY_{n+1}-B)(Y_nY_{n-1}-B)=A^2Y_n,\eqdef\stre$$ 
where we can again scale $B$ to 1 by redefining the variable $Y$. The singularity patterns of (\stre) are $\{0,\infty,\infty,0\}$ and the length-8 cyclic one $\{\infty,0,\infty,\infty,0,\infty,f,f'\}$, which can again be read off directly from (\uvpattii) and (\uvpattiii),  while its deautonomisation is of course $\log A_n=\alpha n+\beta$, (with $B=1$) [\mulkmt]. The invariant of (\stre) is
$$K={(Y_nY_{n-1}-B)(Y_n+Y_{n-1})+A^2\over Y_nY_{n-1}-B}\eqdef\scua$$
As expected, starting from (\scua) we find again the result obtained above, namely equation (\esex). Let us point out here that the asymmetric equation (\miu) is just the Miura relation between (\poct) and (\stre), a result first obtained in [\refdef\japan].

\bigskip
6. {\scap Conclusions}
\medskip

In this paper we addressed the question of the systematic construction of discrete Painlev\'e equations in the degeneration cascade of the affine Weyl group E$_8^{(1)}$. Our starting point was the list of discrete Painlev\'e equations in E$_8^{(1)}$ derived in trihomographic form in previous works of ours. In order to illustrate our method, while keeping the paper at a manageable length, we chose to work with three systems we selected among a dozen candidates. The first step in our method was to consider the autonomous limit of a given equation and, using the results of [\multi], to write the simplest possible mapping in the cascade as well as its invariant. Next, introducing homographic transformations, we obtained all possible canonical forms of the said invariant, the corresponding mapping ensuing automatically. To check that our approach does populate the degeneration cascade without omissions, we compared our results to the classification of singular fibres in [\pers], obtaining a perfect agreement. 

Our method is an alternative to that of Carstea, Dzhamay and Takenawa, who explicitly calculate all singular fibres for a given QRT mapping and then proceed to its degeneration (without however taking into account any possible additive reductions that might arise for special parameter values in the mapping). 
Another advantage of our method is that once a mapping is obtained in canonical form, its deautonomisation, leading to a discrete Painlev\'e equation, is quite straightforward. As it turned out, in most cases presented here this last step was not even necessary since we could find a relation to a previously derived result of ours. Still some new discrete Painlev\'e equations did make their appearance.

This is a first exploratory work, intended to showcase our method and the extreme usefulness of the classification of the QRT canonical forms. We intend to return to our study of the degeneration cascade of E$_8^{(1)}$ in a future work of ours. 

\bigskip
{\scap Acknowledgements}
\medskip

RW would like to acknowledge support from the Japan Society for the Promotion of Science (JSPS), through the the JSPS grant: KAKENHI grant number 15K04893. 

\bigskip
{\scap References}
\medskip

 [\physrep] A. Ramani, B. Grammaticos and T. Bountis, Physics Reports 180 (1989) 159.

 [\sincon] B. Grammaticos, A. Ramani and V. Papageorgiou, Phys. Rev. Lett. 67 (1991) 1825.

 [\hiv] J. Hietarinta and C-M. Viallet, Phys. Rev. Lett. 81, (1998) 325.

 [\redemp] A. Ramani, B. Grammaticos, R. Willox, T. Mase and M. Kanki, J. Phys. A 48 (2015) 11FT02.

 [\takenawa] T. Takenawa, J. Phys. A 34 (2001) L95.

 [\dillerfavre] J. Diller and C. Favre, Am. J. Math. 123 (2001) 1135.

 [\redeem] B. Grammaticos, A. Ramani, R. Willox, T. Mase and J. Satsuma, Physica D 313 (2015) 11.

 [\cdt] A.S. Carstea, A. Dzhamay and T. Takenawa, {\it Fiber-dependent deautonomization of integrable 2D mappings 
and discrete Painlev\'e equations}, preprint (2017) arXiv:1702.04907.

 [\qrt] G.R.W. Quispel, J.A.G. Roberts and C.J. Thompson, Physica D34 (1989) 183.

 [\sakai] H. Sakai, Comm. Math. Phys. 220 (2001) 165.

 [\late] T. Mase, R. Willox, B. Grammaticos and   A. Ramani, Proc. Roy. Soc. A 471 (2015) 20140956.

 [\halb]   R.G. Halburd, Proc. R. Soc. A 473 (2017) 20160831.

 [\expres] A. Ramani, B. Grammaticos, R. Willox and T. Mase, J. Phys. A 50 (2017) 185203.

 [\eight] A. Ramani and B. Grammaticos, J. Phys. A 48 (2015) 355204.

 [\otonom] A. Ramani, S. Carstea, B. Grammaticos and Y. Ohta, Physica A  305 (2002) 437.

 [\jmp] B. Grammaticos and A. Ramani, J. Math. Phys. 56 (2015) 083507.

 [\multi] B. Grammaticos and A. Ramani, J. Phys. A 48 (2015) 16FT02.

 [\mulkmt] B. Grammaticos, A. Ramani, K.M. Tamizhmani, T. Tamizhmani and J. Satsuma, J. Math. Phys. 57 (2016) 043506.

 [\maxim] A. Ramani and B. Grammaticos, J. Phys. A 47 (2014) 385201.

 [\byrd] P.F. Byrd and M.D. Friedman, {\sl Handbook of Elliptic Integrals for engineers and scientists}, Springer-Verlag Berlin, (1971).

 [\seven] A. Ramani and B. Grammaticos, J. Phys. A FT 47 (2014) 192001.

 [\ancil] A. Ramani and B. Grammaticos, J. Phys. A 50 (2017) 055204.

 [\veselov] A.P. Veselov, Russian Math. Surveys 46 (1991) 1.

 [\tsuda] T. Tsuda, J. Phys. A 37 (2004) 2721.

 [\os] K. Oguiso and T. Shioda, Comm. Math. Univ. Sancti Pauli 40 (1991) 83.

 [\pers] U. Persson, Math. Z. 205 (1990) 1.

 [\mir] R. Miranda, Math. Z. 205 (1990) 191.

 [\mase] T. Mase, {\it Studies on spaces of initial conditions for nonautonomous mappings of the plane}, preprint (2017) arXiv:1702.05884.

 [\addkmt] B. Grammaticos, A. Ramani, K.M. Tamizhmani, T. Tamizhmani and J. Satsuma, J. Math. Phys. 55 (2014) 053503.

 [\capel] A. Ramani and B. Grammaticos, Physica A 228 (1996) 160.

 [\sequel] A. Ramani, R. Willox, B. Grammaticos, A.S. Carstea and J. Satsuma, Physica A 347 (2005) 1.

 [\duist] J. J. Duistermaat, {\it Discrete Integrable Systems: QRT maps and Elliptic Surfaces} (Springer-Verlag, New York, 2010)

 [\mickens] B. Grammaticos, A. Ramani, J. Satsuma and R. Willox, J. Math. Phys. 53 (2012) 023506.

\end{document}